# Anomalous Reflection Under Ambient Sunlight: Accessing In-Plane Radiation Pressure for Solar Sailing

Tom Joly-Jehenne[1,2] and Artur R. Davoyan[1,*]

[1]Department of Mechanical and Aerospace Engineering, University of California, Los Angeles, CA, 90095, USA
[2]Department of Physics, Ecole Normale Supérieure Paris-Saclay, 4 Avenue des Sciences, Gif-sur-Yvette, 91190, France
davoyan@seas.ucla.edu

**Abstract:** Harnessing solar radiation pressure is key to transforming space exploration with multiple low cost sunlight propelled spacecraft to outer reaches of space. By controlling the direction of sunlight momentum transfer new missions and better maneuvering in space can be accessed. Here, we discuss design principles for taming in-plane radiation pressure under ambient sunlight. We propose and study theoretically ultra-wideband polarization insensitive metasurfaces for anomalous light reflection. Our design based on segmented tapered patch nanoantenna arrays allows reflection of >60% into one diffraction orders over a 400 nm band across larger part of the solar spectrum. Owing to a wideband nature and polarization insensitivity, our structures convert incident radiation into in-plane radiation pressure force with almost 30% efficiency. We discuss applications of our design to controlling solar sail spin. Beyond solar sailing, we envision that such anomalous metasurfaces for ambient sunlight will find use in solar concentration, spectrum splitting, and solar fuels.

**Main text:**

Solar sailing is of great promise for future space exploration [1-3]. Indeed, solar sails make use of radiation pressure for propulsion and, therefore, are not constrained by inherent limitations of conventional rockets [3, 4]. In particular, solar sails can reach orbits and destinations that are inaccessible to present day spacecraft [1, 5], enabling a range of breakthrough science missions. Such missions as solar polar imaging [3, 5, 6], non-Keplerian station keeping [7, 8], artificial Lagrange points and halo orbits for early space weather warning [9, 10], as well as fast transit probes to outer planets and interstellar medium [2, 3, 5] have been envisioned recently.

Present day solar sails are made of thin specularly reflective films (typically, aluminum coated polyamides [3, 10-12]). Specular reflection implies that radiation pressure force is always directed along the local normal to the sail surface and is independent of the angle of sunlight incidence (assuming negligible sunlight absorption) [3]. Recently it was proposed that by scattering light in nonspecular directions, for example with the use of diffraction gratings [13-16] or phase gradient metasurfaces [17, 18], incident radiation can impart momentum in the plane of the sail. It was further suggested that accessing such lateral radiation pressure forces is of great promise for sail controls [3, 18-20] and more efficient mission design [21, 22]. The ability of imparting in-plane momentum under laser illumination was also examined and verified in several of recent experiments [23, 24]. Geometries that can harness in-plane momentum transfer are also of great importance in the context of laser sailing [25, 26], where approaches for passive self-stabilized laser beam riding are needed [26].

Harnessing in-plane radiation pressure forces with diffraction gratings has been extensively studied in [13-16, 23, 24]. Symmetric and blazed transmission-based diffraction gratings were analyzed both theoretically and experimentally, and conditions that optimize in-plane momentum transfer were outlined. For symmetric gratings lateral radiation pressure force is maximized for oblique incidence angles [13, 23]. However, the need for an oblique incidence challenges implementation of symmetric gratings in the context of the solar sailing (as it would require sails surfaces oriented at a relatively large angle with respect to sunline). Transmission blazed gratings can operate under normal incidence and therefore circumvent this requirement [14-16, 24]. Radiation pressure transfer under broadband illumination for such gratings was studied in [14, 15]. It was shown that overall in-plane momentum transfer efficiency, $\eta_\parallel$, grows as a function of grating period, $\Lambda$, and for a theoretical idealistic scenario it reaches $\eta_\parallel > 90\%$ for $\Lambda > 50\ \mu m$, assuming linearly polarized light [14] (efficiency is defined as $\eta_\parallel = \frac{F_\parallel}{P/c}$, where $F_\parallel$ is the net in-plane radiation pressure force and $P/c$ is "incident" radiation pressure, $P$ is impinging power, and $c$ is the speed of light). More realistic full wave simulations showed that the efficiency of such transmission blazed gratings is limited to $\eta_\parallel \sim 30\%$ even at an oblique incidence (a hypothetical case of a high index structure and linearly polarized incident light were assumed) [15]. Furthermore, as the blazed grating period grows its areal density increases making such gratings less suitable for solar sailing (e.g., a $20^o$ blaze grating with a $50\ \mu m$ period would correspond to an average $\sim 10\ \mu m$ thick layer; compare with $2.5\ \mu m$ thick sail material used in NEA Scout mission [12]). Transmission diffraction gratings with an asymmetric cell may help designing thin film metasurface structures [16, 27, 28]. A recent full wave study of such asymmetric thin film silicon nitride gratings, however, revealed only a modest in-plane momentum transfer efficiency [16]. For completeness, it is worth mentioning use of reflection blazed gratings [19] for in-plane momentum transfer. Measurements under a normal incidence indicate about 4% efficiency [19]. Although higher efficiency may be reached with a better design, operating at a normal incidence angle for reflection blazed grating is inherently suboptimal (maximum efficiency is reached at a Littrow angle [29]). In addition, both transmission and reflection blazed gratings suffer from effects of shadowing [15, 19] and are sensitive to the angle of incidence [29] and light polarization.

Phase gradient metasurfaces provide an alternative approach to the design of optical systems [30, 31]. In particular, thin metasurface films allow a highly agile manipulation of optical fields, including polarization [32], phase [33], spectral [34] and direction-dependent [35] selectivity. The ability of metasurfaces for near-arbitrary control over light scattering is also of great interest for radiation pressure management in the context of laser [36] and solar sailing [16-18]. While phase gradient metasurfaces have been proposed for solar sailing [17-18], detailed design of such systems and study of their performance under broadband unpolarized sunlight is yet to be performed. Here, inspired by prior works on anomalous light reflection with plasmonic metasurface reflect arrays [37-41], we examine theoretically design principles for efficient in-plane momentum transfer under ambient sunlight. We show that metasurface reflect arrays based on segmented tapered patches (Fig. 1a) yield a *nearly polarization independent anomalous reflection over a 400 nm band across visible and near-infrared*. Such broadband performance allows inducing efficient lateral solar radiation pressure force with in-plane momentum coupling coefficient under unpolarized sunlight reaching $C_{\parallel} \simeq 0.65\ nN/W$ ($\eta_{\parallel} \sim 20\%$) for normal incidence and $C_{\parallel} \simeq 0.92\ nN/W$ ($\eta_{\parallel} \sim 28\%$) for $30^o$ incidence angle. We further discuss role of materials and study an example application of such metasurfaces to control solar sail dynamics.

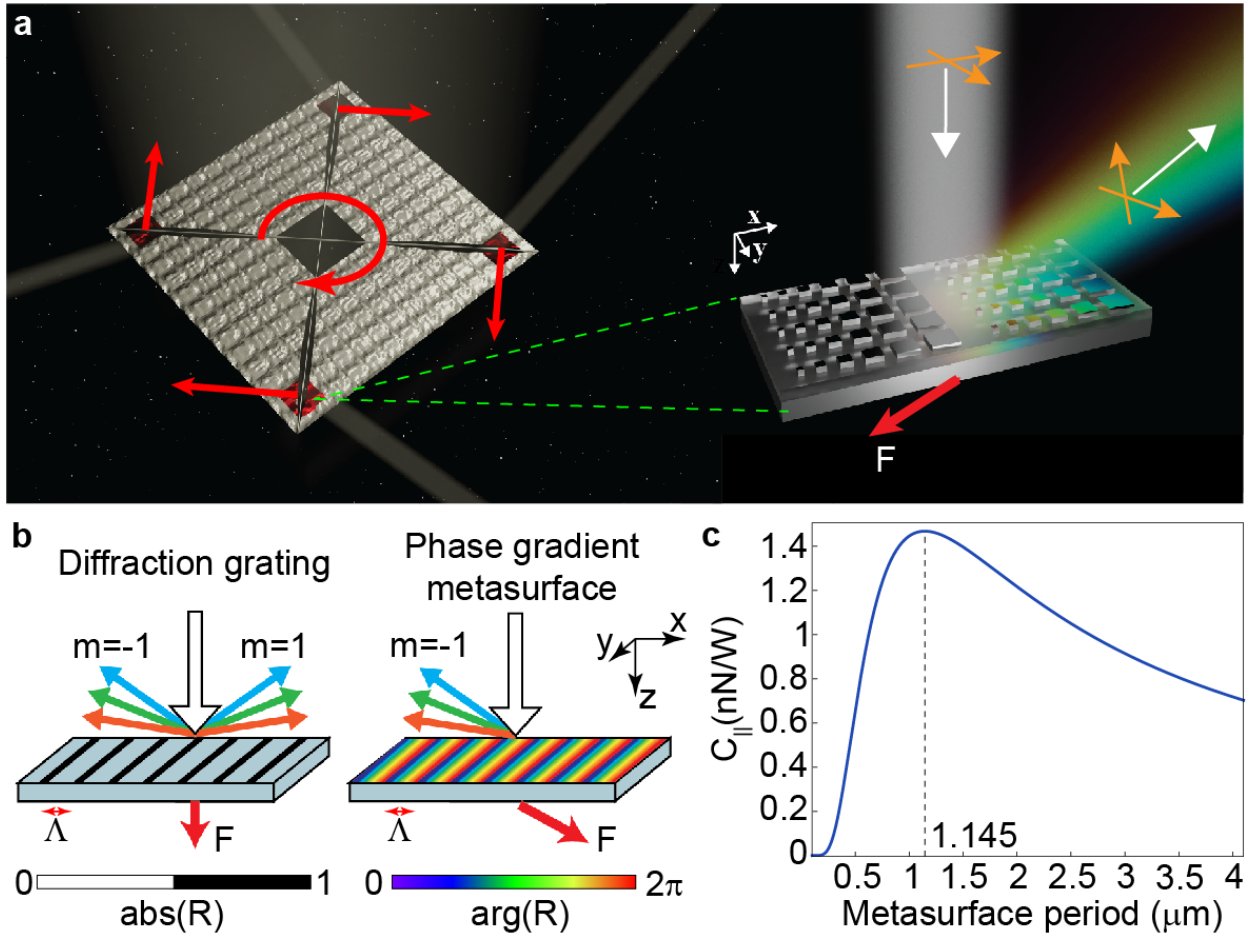

*Fig. 1. Broadband polarization independent anomalously reflecting metasurface for solar sail control. (a) Schematic illustration of a solar sail with metasurfaces added at the corners of the sail for controlling sail attitude (roll around the z axis). Zoomed-in is a conceptual illustration of a broadband polarization independent reflect array metasurface structure for inducing lateral radiation pressure force, proposed and studied in this work. (b) Comparison between a regular symmetric reflection diffraction grating and phase gradient reflect array metasurface. The 2π phase gradient across the period can lead to a selective reflection into one of the diffraction orders and, as a result, to a net in-plane momentum transfer. (c) Calculated theoretical momentum coupling coefficient as a function of the metasurface period, assuming an ideal 100% reflectivity in the $m = -1$ diffraction order only.*

We begin our study by revisiting basic principles of solar radiation pressure in-plane momentum transfer with phase gradient metasurfaces [33]. For the sake of simplicity of discussion, consider a solar sail made of a one dimensional (1D) periodic reflect array grating with period $\Lambda$, Fig. 1b. Without any phase gradient across the unit cell ($\nabla\Phi = 0$), under normal incidence such a symmetric periodic reflect array grating will scatter light into diffraction orders, which are defined by a diffraction relation [23, 41] $\frac{2\pi}{\lambda}\sin\theta_{\lambda,m} = m\frac{2\pi}{\Lambda}$, where $\theta_{\lambda,m}$ are directions of wave reflection at wavelengths $\lambda$ with $m = \cdots -1, 0, 1 \cdots$ denoting diffraction orders. Waves reflected into nonzero diffraction orders (i.e., $|m| > 0$) impart radiation pressure force in the plane of the sail: $F_{\parallel,m} = -\frac{A}{c}\frac{(1AU)^2}{r^2}\int d\lambda I_{sun}(\lambda) R_m(\lambda)\sin\theta_{\lambda,m}$, where $I_{sun}(\lambda)$ is the AM0 solar spectral power density, which in this work for sake simplicity of analysis we approximate as a black body spectrum at temperature $5800\,K$, $R_m(\lambda)$ is the reflection spectral amplitude in the $m$-th order, $A$ is the metasurface area, $r$ is the distance from the sun in astronomical units (AU), and $c$ is the speed of light. The overall lateral force is then computed by summing the contribution from all forces induced by each diffraction order:

$$|F_\parallel| = \left|\sum_m F_{\parallel,m}\right| = \frac{A}{c}\frac{(1AU)^2}{r^2}\int I_{sun}(\lambda)\sum_{m=-\infty}^{\infty} R_m(\lambda)\sin(\theta_{\lambda,m})d\lambda. \quad (1)$$

For a symmetric array (i.e., when $\nabla\Phi = 0$) momentum transfer due to light reflected in $\pm m$ diffraction orders is equivalent, i.e., $|F_{\parallel,-m}| \equiv |F_{\parallel,m}|$, as a result the net in-plane momentum is zero (i.e., $|F_\parallel| = 0$). With a phase gradient of $2\pi$ across array unit cell, conditions for an anomalous reflection can be reached [37, 38, 40], and the array behaves as an anomalously reflecting metasurface [33, 37, 38]. In this case, incident light is predominantly reflected into one of the nonzero diffraction orders (typically with $|m| = 1$), resulting in a net lateral radiation pressure force ($|F_\parallel| > 0$). The net lateral force, $F_\parallel$, evidently, depends on the metasurface period $\Lambda$. For example, only wavelengths with $m\lambda < \Lambda$ can be diffracted into the $m$-th diffraction order, which means that for $\lambda > \Lambda$, only specular reflection (i.e., diffraction into $m = 0$ order) pertains even for phase gradient metasurfaces. As such wavelengths with $\lambda > \Lambda$ do not contribute to the in-plane momentum transfer (normal incidence is assumed).

Assuming an idealistic scenario of perfect polarization independent anomalous reflection into the $m = -1$ diffraction order only (i.e., for $\lambda < \Lambda$, $R_m(\lambda) = \{1\ if\ m = -1;\ 0\ if\ m \neq -1\}$) across entire spectrum, we can put an upper bound on the lateral radiation pressure force:

$$|F_\parallel|_{max} = \frac{A}{c}\frac{(1AU)^2}{r^2}\int_0^\Lambda I_{sun}(\lambda)\sin(\theta_{\lambda,-1})d\lambda, \quad (2)$$

To assess the efficiency of in-plane momentum transfer we introduce a related in-plane momentum coupling coefficient: $C_\parallel = \frac{F_\parallel}{P}$, where $P = A\frac{(1AU)^2}{r^2}\int I_{sun}(\lambda)d\lambda$ is the total impinging solar power. Note that the momentum coupling coefficient is related with the momentum transfer efficiency [14] as $\eta_\parallel = c \times C_\parallel$. In Fig. 1c we plot the in-plane momentum coupling coefficient for such an idealistic scenario (i.e., Eq. (2)). We find that momentum coupling coefficient reaches maximum of $C_\parallel \simeq 0.145 \frac{nN}{W}$ at $\Lambda_0 \simeq 1\mu m$ metasurface period ($\Lambda_0 \simeq 1.145\ \mu m$ to be exact), which corresponds to $\eta_\parallel \simeq 0.435$. For metasurfaces with a smaller grating period (i.e., $\Lambda < \Lambda_0$) a significant fraction of solar radiation has wavelengths longer than the grating period. Solar radiation at these wavelengths is reflected specularly without contributing to in-plane momentum transfer. On the contrary, larger metasurface period (i.e., $\Lambda > \Lambda_0$), while allows accessing a broader spectral range, implies anomalous reflection into smaller angles $\theta_{\lambda,-1}$ (specifically, $\sin(\theta_{\lambda,-1}) = -\frac{\lambda}{\Lambda} \to 0$ for $\lambda \ll \Lambda$). As a result for metasurfaces with $\Lambda > \Lambda_0$ the net in-plane momentum transfer is reduced (see Eq. (1)). Fig. 1c and Eq. (2) provide an overall guideline to metasurface design.

Anomalous light reflection with phase gradient metasurfaces has received a broad attention in recent years [30, 31, 33, 37, 38]. In the context of solar sailing of particular interest are metasurface reflect arrays [41] that can reflect incident light anomalously [18, 37, 38]. At the core of such metasurfaces are unit cells in which $2\pi$ phase gradient across the unit cell is attained [37]. Previously it was shown that plasmonic patch nano-antennas allow a wide tunability of reflection phase and accessing the full $2\pi$ phase band [37, 39]. A range of reflect array metasurfaces based on plasmonic patched nano-antennas has been proposed [37, 39, 41], including high efficiency surfaces for anomalous light reflection [37]. However, the bandwidth over which anomalous reflection is achieved is rather limited in these systems (i.e., not suitable for broadband solar spectrum management). In earlier works a unit cell of a phase gradient reflect array metasurface comprised of a uniform array with a discrete set of rectangular plasmonic patched nano-antennas of varying length, which approximated the required linear $2\pi$ phase gradient across [37] (see also Fig. 3a). In [38] such a discrete array was replaced with a single tapered patch nanoantenna to approximate continuous $2\pi$ phase variation across the unit cell. Importantly, the design allowed achieving a broadband operation across the visible spectrum [38, 40]. However, in both of the designs (i.e., discrete patched antenna array and single tapered patch) the operation remains polarization sensitive. That is, anomalous light reflection is achieved predominantly for one of the polarizations. Such polarization sensitivity limits the application of these structures for solar sailing under broadband unpolarized sunlight. Here, we utilize advantages of both approaches to design a broadband polarization independent anomalously reflecting metasurface reflect array, Fig. 1a.

As a starting point of our metasurface design we begin with studying in-plane momentum transfer by an anomalously reflecting metasurface based on a single tapered patch antenna. Unit cell of such a metasurface is schematically shown in Fig. 2a, and is largely inspired by [38]. For such a structure light polarized along the shorter axis direction (i.e., along $y$ axis) resonantly excites the patch antenna resulting in an anomalous reflection [38]. We study silver (Ag) and aluminum (Al) based plasmonic metasurfaces. In both cases we assume $30\ nm$ thick antenna patch separated by a thin alumina ($Al_2O_3$) spacer layer from the ground metallic plane. While Ag is low loss in the visible and near infrared [42, 43], Al allows accessing ultraviolet part of the spectrum [44]. In addition, surface plasmon cut-off frequency, $\omega_{sp}$, at the metal – dielectric interface ($\omega_{sp} = \omega_p/\sqrt{1+\varepsilon}$) extends to a higher frequency for Al as compared to Ag owing to its higher bulk plasma frequency ($\omega_p \approx 15\ eV$ for Al against $\omega_p \approx 9\ eV$ for Ag [43]). Higher frequency cut-off, in turn, translates to accessing shorter wavelengths with Al plasmonic patch antennas. The unit cell is periodically laid out in a two dimensional array. Consistent with our analysis of the idealistic scenario (see Fig. 1c), we choose the period in the direction of phase gradient (i.e., in the direction patch taper along the $x$ axis) as $\Lambda = 1\ \mu m$. It is along this direction (i.e., $x$ axis, Fig. 2a) that anomalous light reflection is expected due to phase gradient [38]. The transverse period along the other direction, $w$, is chosen as $w \ll 1\ \mu m$ to avoid spurious diffraction along the $y$ axis for a major part of the solar spectrum (i.e., wavelengths with $\lambda > w$ experience no diffraction along the $y$ axis).

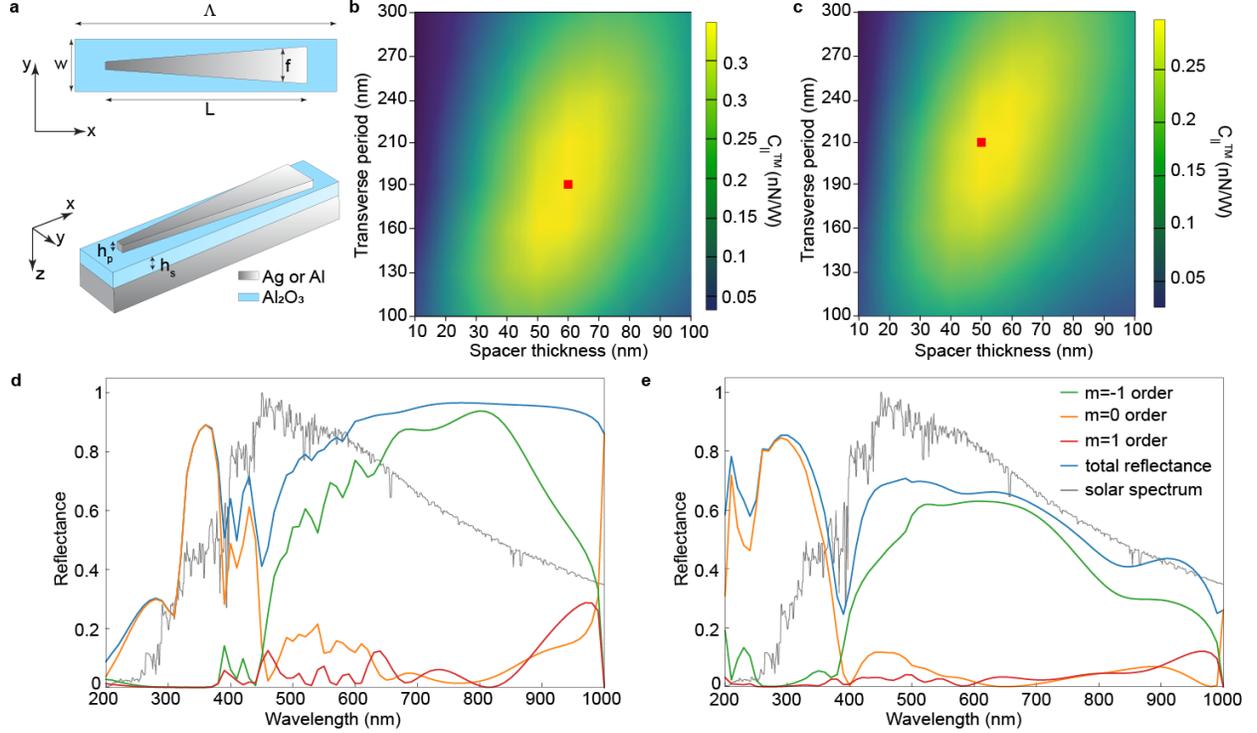

*Fig. 2. Understanding momentum coupling coefficient. (a) Schematic illustration of a metasurface unit cell geometry. Here $L = 800\ nm$, $\Lambda = 1000\ nm$, $h_p = 30\ nm$, and $f$ varies from $0.15w$ to $0.75w$. (b) Calculated in-plane momentum coupling coefficient, $C_\parallel^{TM}$, as a function of the spacer thickness, $h_s$, and transverse period, $w$, for an Ag based metasurface. (c) same as in panel (b) but for an Al based metasurface. In (b) and (c) markers denote corresponding maxima of $C_\parallel^{TM}$. (d) Reflectance spectra into $m = -1, 0, 1$ diffraction orders for an Ag based metasurface at $w = 190\ nm$ and $h_s = 60\ nm$. (e) Reflectance spectra into $m = -1, 0, 1$ diffraction orders for an Al based metasurface at $w = 210\ nm$ and $h_s = 50\ nm$. In (d) and (e) Total reflection and AM0 spectra are also shown for comparison.*

In Figs. 2b and 2c we plot the variation of the in-plane momentum coupling coefficient, $C_\parallel^{TM}$, as a function of spacer thickness, $h_s$, and transverse period, $w$ (see also Fig. 2a) for a normally incident TM polarized light (i.e., with electric field $\boldsymbol{E} = (0, E, 0)$) for Ag and Al based metasurfaces, respectively. Specifically, we calculate numerically reflectance spectra, $R_m(\lambda)$, into $m = -2, -1, 0, 1, 2$ diffraction orders and then estimate momentum coupling coefficient as: $C_\parallel^{TM} =$

$\frac{1}{cI_0} \sum_m \int \frac{1}{2} I_{sun}(\lambda) R_m(\lambda) \sin \theta_{\lambda,m} \, d\lambda$, where $I_0 \simeq 1360 \, W/m^2$ is AM0 solar irradiance, the factor $1/2$ is introduced to account for incident linear polarization. A similar expression can be written for an incident TE polarization (i.e., $\mathbf{E} = (E, 0, 0)$). The total in-plane momentum coupling coefficient for unpolarized sunlight is then found as $C_{\parallel} = (C_{\parallel}^{TM} + C_{\parallel}^{TE})$. For both Ag and Al based metasurfaces (Fig. 2b and Fig. 2c) we observe that as the parameters of the unit cell vary, in-plane momentum coupling coefficient reaches maximum. In case of an Ag based metasurface the optimum corresponds to $C_{\parallel}^{TM} \simeq 0.4 \frac{nN}{W}$ at $w = 190 \, nm$ and $h_s = 60 \, nm$, whereas for Al based metasurface optimum $C_{\parallel}^{TM} \simeq 0.29 \frac{nN}{W}$ at $w = 210 \, nm$ and $h_s = 50 \, nm$. Interestingly, despite difference in material properties, in both studied cases the optimum performance is found for a near similar spacer layer thickness and transverse period.

Moving forward, to better understand the observed dynamics, we examine spectrally resolved reflectance $R_m(\lambda)$ at respective maxima of in-plane momentum coupling coefficient for Ag (Fig. 2d) and Al (Fig. 2e) based metasurfaces, respectively. Firstly, as expected [38], both Ag and Al structures reflect light anomalously into $m = -1$ diffraction order across a very broad spectral range. Secondly, no anomalous reflection for $\lambda > \Lambda = 1000 \, \mu m$ is observed, as predicted. Further, we observe that overall Ag based metasurface is more efficient in reflecting light as compared to Al based metasurface, which we attribute to a lower loss in Ag across visible and near infrared bands [42, 43]. Specifically, anomalous reflection into $m = -1$ order for Ag based metasurface reaches almost $R_{-1} \simeq 0.94$ at $800 \, nm$ ($R_{-1} > 0.8$ across the $700 \, nm - 900 \, nm$ band), whereas for Al based metasurface the maximum is $R_{-1} < 0.6$. At the same time, for Ag based metasurface anomalous reflection is not observed below $\lambda \simeq 450 \, nm$, which corresponds to a plasmon-polariton cut-off wavelength for Ag/Al$_2$O$_3$ interface (potential use of lower index spacer layers, e.g., MgF$_2$, is expected to extend cut-off slightly further into UV). In contrast, while Al based metasurface exhibits lower overall reflectance, it allows reflecting light anomalously down to $380 \, nm$, therefore covering most of the solar spectrum. We note that, as both Ag and Al based metasurfaces reflect light anomalously predominantly for TM polarization, the overall efficiency for an unpolarized sunlight is limited. For comparison, for an Ag based metasurface we have examined reflectance under TE polarization as well. For this polarization a slight imbalance in reflection between $m = 1$ and $m = -1$ orders is also observed, with $m = 1$ slightly dominating (see also Fig. 3c). As a result, a small in-pane momentum transfer in an opposite direction (as compared to the TM case) is attained under TE polarization (see also Fig. 3b). Calculated in-plane momentum coupling coefficient yields $C_{\parallel}^{TE} \approx -0.127 \frac{nN}{W}$. Due to opposite directions of in-plane momentum transfer for incident TE and TM polarizations, the overall in-plane momentum coupling coefficient is relatively low: $C_{\parallel} = (C_{\parallel}^{TM} + C_{\parallel}^{TE}) = 0.273 \frac{nN}{W}$, i.e., $< 19\%$ of the theoretical maximum studied in Fig. 1c.

Moving forward, we examine strategies that enable polarization independent anomalous reflection. The structure studied in Fig. 2a is inspired by the woks [38, 40], and as mentioned, is tailored to operate predominantly under incident TM polarization. Light polarized along the shorter, transverse period direction, "sweeps" continuous ~$2\pi$ phase variation along the tapered patch antenna (Fig. 3a). Desired $2\pi$ phase gradient across metasurface unit cell can also be well approximated by a discrete array of plasmonic patch nanoantennas of varying length [37]. In this case metasurface unit cell contains an array of nanoantennas of equal width but varying length, Fig. 3a [37]. Light polarized along nanoantenna length (i.e., TM polarization) excites dipolar modes (along $y$-axis) with a relative phase of reflected light defined by the element length. At the same time incident light polarized along nanoantenna width (i.e., TE polarization) excites dipolar modes along the $x$-axis. However, as the width of the patches is fixed, no substantial relative phase variation is attained for this polarization. To obtain desired phase variation for both TE and TM polarizations over a broad spectral range, we combine design principles of a single tapered patch antenna and that of an array, Fig. 3a. Specifically, we consider an array of patches where both the width and length of patches are varied, see Fig. 3a, so as phase gradient is achieved for both incident polarizations. This design is achieved by effectively segmenting a single tapered patch antenna into an array of tapered patches (Fig. 3a). In this case tapered shape of individual patch segments is further expected to preserve desired broad band operation. We note that polarization independent reflect arrays based on circular patches of varying radius have been considered before for the mid-infrared band [39]. However, their properties, such as broadband anomalous light reflection across visible and infrared bands, and especially use for in-plane momentum transfer studied here, remain unknown.

To examine the effectiveness of segmented tapered patch design, we study in-plane momentum coupling coefficient for both TE and TM polarizations under normal incidence, following the procedure outlined earlier. For the purpose of demonstration, we consider an Ag based metasurface (studied in Figs. 2b and 2d for TM polarized excitation). In Fig. 3b we plot in-plane momentum coupling coefficient as a function of the number of segments calculated for TM and TE incident polarizations, respectively. We observe that for a segmented design the in-plane momentum coupling coefficient for TM polarization is smaller than that for a single tapered patch ($C_\parallel^{TM} \simeq 0.4\ nN/W$ for a single tapered patch and $C_\parallel^{TM} \simeq 0.3\ nN/W$ for 6 tapered patch segments per unit cell; i.e. ~25% decrease). At the same time, segmenting a single tapered patch into an array, allows boosting in-plane momentum coupling coefficient for the TE polarization, $C_\parallel^{TE}$. In this case $C_\parallel^{TE}$ has a clearly defined maximum as a function of a number of segments. Specifically, the maximum efficiency for TE polarization is obtained for 6 patch segments per unit cell and reaches $C_\parallel^{TE} \simeq 0.25\ nN/W$. The overall in-plane momentum coupling coefficient for unpolarized light for 6 segments is then $C_\parallel \simeq 0.65\ nm/W$ (compare with $C_\parallel \simeq 0.273\ nN/W$ for a single tapered patch studied in Fig. 2b). As the number of patch segments is further increased (Figs. 3b and 3c), the spacing between patches decreases resulting in cross coupling of individual patches for TE polarization, which decreases overall performance, as is seen for the case of 8 patch segments.

To better understand the evolution of the in-plane momentum coupling coefficient with the number of segments for TE polarization, in Fig. 3c we study spectrally resolved ratio of $m = -1$ and $m = +1$ diffraction orders. Specifically, we plot $\log_{10}(R_{-1}(\lambda)/R_1(\lambda))$ for several different patch segment configurations. Anomalous light reflection into $m = -1$ order corresponds to $|R_{-1}(\lambda)/R_1(\lambda)| \gg 1$. For a single tapered patch, studied in Fig. 2a, we observe that $\frac{R_{-1}(\lambda)}{R_1(\lambda)} \leq 1$ across entire visible spectrum. Notably, in this case diffraction into $m = +1$ order slightly dominates, resulting in a nonzero in-plane momentum transfer for TE polarization even for a single tapered patch design ($C_\parallel^{TE} \simeq -0.127\ nN/W$). For 3 segments the strength of diffraction into $m = +1$ and $m = -1$ orders varies across the visible spectrum, resulting in a vanishingly small spectrally averaged in-plane momentum transfer. For 6 segments diffraction into $m = -1$ order dominates and a very strong asymmetry in light reflection is observed. A respective spectrally resolved reflectance into different diffraction order under incident TE polarization for 6 segmented tapered patches in plotted in Fig. 3d. Over 50% reflection into $m = -1$ diffraction order is seen between $\sim 600\ nm$ and $950\ nm$, with nearly complete reflection into $m = -1$ order at $\lambda \simeq 860\ nm$. At the same time between 400 nm and 600 nm the structure exhibits a relatively strong parasitic absorption. Such absorption may lead to an excessive heating for missions close to the sun and may require extra thermal management [3]. In this work we are interested in laying out basic principles based on relatively simple intuitive approaches. We anticipate that with the use advanced inverse design methods [45] more complex and more efficient structures, which are particularly suited for a given solar sail mission application can be created.

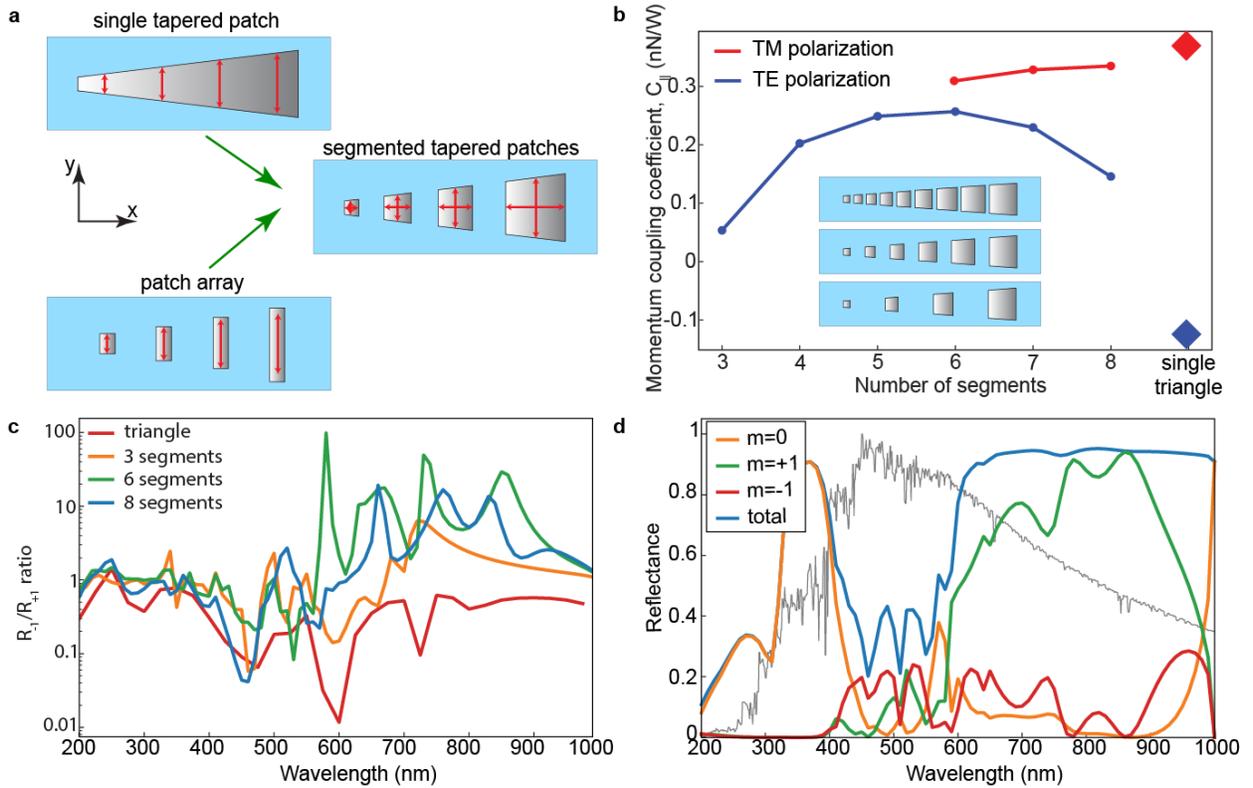

***Fig. 3**. **Polarization independent design for in-plane momentum transfer.** (a) Schematic illustration of a design approach to achieving broadband polarization independent anomalous reflection. (b) In-plane momentum coupling coefficient as a function of a number of segmented*

*tapered patches for TM and TE polarizations for an Ag based metasurface design. Momentum coupling coefficient for a single tapered patch (i.e., a single triangular antennae) is shown for comparison. Inset shows several examples of segmented tapered patches metasurface unit cells. (c) Spectrally resolved ratio of reflectance into $m = -1$ and $m = +1$ diffraction orders for several differed designs. (d) Spectra of reflection into m=-1,0,1 diffraction orders for an Ag based metasurface for a design with 6 segmented tapered patches. Total reflectance and solar spectrum are shown for comparison.*

Next, for practical solar sail applications it is important to understand variation of the in-plane momentum transfer as a function of the angle of incidence. Indeed, owing to sail deformations and variation of the sail orientation over the course the mission, the local angle of incidence may vary [3, 12]. Considering as an example an Ag based metasurface design with 6 patch segments (Fig. 3), we study in-plane momentum transfer as a function of the incidence angle. In Fig. 4a we plot in-plane momentum coupling coefficient for an polarized light, $C_\parallel$, as a function of incident direction expressed through azimuthal, $\phi$, and polar, $\psi$, angles in a spherical coordinate system (Fig, 4b shows schematically the geometry). We observe that for incidence angles in the range $-90^o < \phi < 90^o$ and $\psi < 10^o$ the metasurface exhibits nearly steady performance. For larger polar angles $\psi > 10^o$ the momentum coupling coefficient gradually changes its value from positive to negative values as the azimuthal angle, $\phi$, is varied from $0^o$ to $180^o$. Change in the sign of the momentum coupling coefficient implies that the direction of lateral radiation pressure force along the $x$ axis changes as a function of the polar angle, $\phi$. We remind that, irrespective of the incidence angle and light polarization, in-plane momentum transfer occurs only along the $x$ axis, i.e., along the phase gradient axis. To further illustrate the variation with the angle of incidence, we plot in-plane momentum coupling coefficient, $C_\parallel$, as a function of the polar angle $\psi$ for several different azimuthal angles, Fig. 4c. For azimuthal angles $|\phi| \leq 45^o$ the momentum coupling coefficient grows as a function of the polar angle reaching local maximum for polar angles in the range $30^o \leq \psi \leq 40^o$. Maximum in-plane momentum coupling coefficient of $C_\parallel \simeq 0.92 \, nN/W$, which is ~63% of the theoretical $C_{\parallel max}$ studied in Fig. 1c, is attained at $\psi \simeq 30^o$ and $\phi = 0^o$. For $\phi = 90^o$ the momentum coupling decreases with the increase of the polar angle eventually approaching $C_\parallel \simeq 0$ for $\psi = 60^o$. On the contrary, for azimuthal angles $135^o \leq \phi \leq 225^o$ $C_\parallel$ decreases and changes its sign as the polar angle grows. Eventually for high polar angles $\psi > 40^o$ the in-plane momentum coefficient reaches a minimum (i.e., maximum lateral radiation pressure force in the $-x$ direction). Specifically, $C_\parallel \simeq -0.95 \, nN/W$ at $\phi = 180^o$ and $\psi = 50^o$.

To better understand the observed dynamics we plot $R_m(\lambda)$ as a function of the diffraction angle $\theta_m$ in the $x - z$ plane (i.e., light diffraction plane) across the $200 \, nm - 1000 \, nm$ spectral band. At normal incidence ($\phi = 0^o$ and $\psi = 0^o$) light is predominantly diffracted into $m = -1$ order, as designed, Fig. 4d. As the polar angle is increased ($\phi = 0^o$ and $\psi = 30^o$) the scattering into $m = -1$ order further increases, Fig. 4e. Now as the azimuthal angle is flipped to $180^o$ ($\phi = 180^o$ and $\psi = 30^o$) diffraction into $m = +1$ order dominates, Fig. 4f, giving raise to $C_\parallel < 0$ (see

Figs. 4a and 4c). Figures 4g and 4h highlight the transition from $m = -1$ to $m = +1$ as the azimuthal angle, $\phi$, varies.

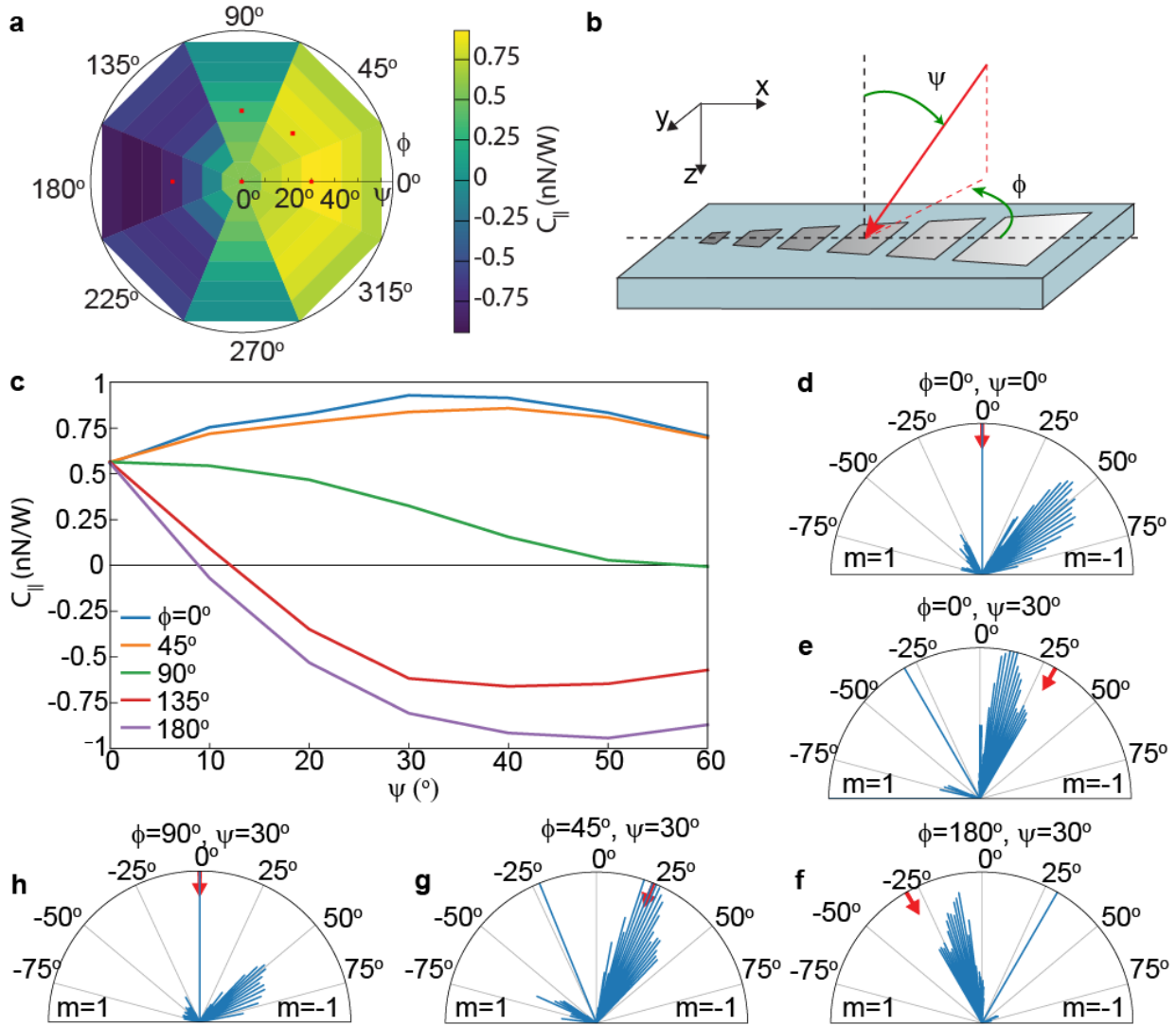

**Fig. 4**. *Angle dependence study. (a) Calculated in-pane momentum coupling coefficient, $C_{\parallel}$, under unpolarized incident light as a function of polar, $\phi$, and azimuthal, $\psi$, angles. (b) Schematic illustration of the studied geometry. (c) Evolution of the in-plane momentum coupling coefficient as a function of polar angle for different azimuthal angles. Light diffraction in $x - z$ plane for $\phi = 0^0$ and $\psi = 0^0$ (d), $\phi = 0^0$ and $\psi = 30^0$ (e), $\phi = 180^0$ and $\psi = 30^0$ (f), $\phi = 45^o$ and $\psi = 30^0$ (g), and $\phi = 90^0$ and $\psi = 30^0$ (h), respectively. In (d-h) arrows denote the direction of incidence as projected onto $x - z$ plane. Markers in (a) correspond to angles studied in (d-h).*

Finalizing our study, we discuss a simple example of sail attitude control with designed anomalously reflecting metasurfaces. We consider a sun-facing solar sail and examine time it takes to rotate the sail around its normal axis (i.e., roll around the z axis, see Fig. 1a). We note that such a maneuver is not possible for a perfectly flat specularly reflective sail, as in case of a flat sail radiation pressure exerts forces only perpendicular to the sail surface. For our analysis we consider a common square sail with diagonal booms [3, 10-12], as is schematically depicted in Fig. 1a. To control sail roll we assume four anomalously reflecting $1\ m^2$ metasurfaces embedded at the corners of the sail [10, 12, 19, 20], as shown in Fig. 1a. Each of the metasurfaces creates torque along the z axis $M_z \simeq \frac{\sqrt{2}}{2} F_{||} L$, where $L$ is the length of the sail side and $F_{||} = C_{||} I_0 \frac{(1AU)^2}{r^2} \times 1\ m^2$ is the lateral radiation pressure force induced by the anomalous metasurfaces. Sail angular acceleration is then found as $\dot{\Omega}_z = \frac{M_z}{I_{zz}}$, where $I_{zz}$ is mass moment of inertia along the z axis and $\Omega_z$ is respective angular velocity. Functionally a square sail spacecraft consists of three main elements: a thin film sail membrane, booms that support the sail membrane, and a spacecraft bus at the center of the sail, which contains all avionics, electronics and scientific instruments [3, 10-14], Fig. 1a. As the spacecraft bus is typically much smaller in size than the sail itself [10], to the first order approximation its contribution to mass moment of inertia can be neglected. In this case mass moment of inertia is expressed as: $I_{zz} \simeq \frac{1}{6} L^2 (\rho_{sail} L^2) + \frac{2}{12} L^2 (\rho_{boom} \sqrt{2} L)$. For the sake of concept demonstration, we assume sail membrane areal density of $\rho_{sail} = 5\ g/m^2$ and booms linear density of $\rho_{boom} = 50\ g/m$ [3, 10]. Time to make a 180° turn is then simply found by integrating angular acceleration twice: $T_{180°} = \sqrt{2\pi \frac{I_{zz}}{M_z}}$. Worth examining this expression in a lmit of a very large sail [3, 10]. In this case mass moment of inertia is dominated by sail membrane contribution, $I_{zz} \to \frac{1}{6} L^2 (\rho_{sail} L^2)$, and, as sail size, $L$, grows, $T \propto rL^{3/2}$. Evidently, time to turn a sail grows linearly with the distance from the sun, $r$, and superlinarly with the sail side length, $L$.

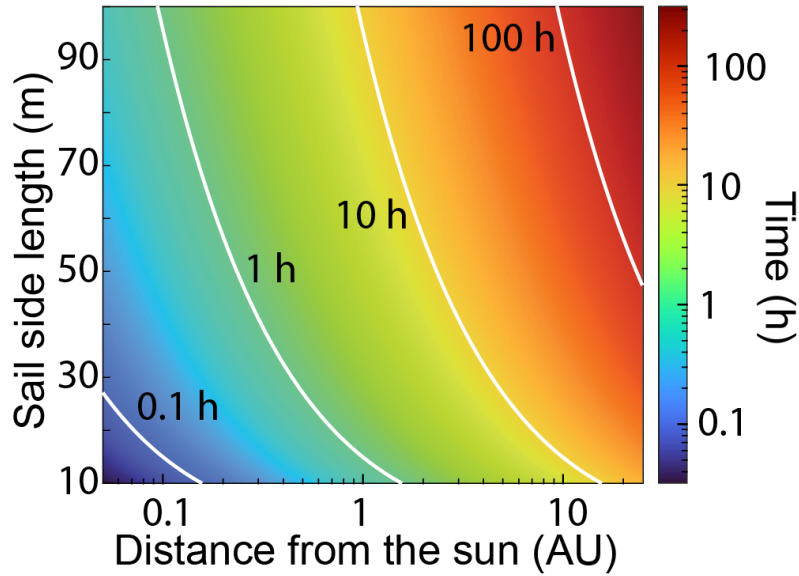

**Fig. 5. Solar sail roll study.** *Calculated time to make a 180° turn as a function of distance from the sun, $r$, and sail side length, $L$. Geometry shown in Fig. 1a is assumed: four $1\ m^2$ anomalously reflecting metasurfaces with in-plane momentum coupling coefficient of $C_{||} = 0.65\ nN/W$ are embedded at the corners of a square sail.*

Assuming in-plane momentum coupling coefficient $C_{||} = 0.65\ nN/W$, i.e., that for a 6 patched segments (Fig. 3b), in Fig. 5 we plot time it takes to make a half turn (i.e., $180^o$ turn) as a function of distance from the sun, $r$, and sail side length, $L$. As expected, close to the sun even for a large size sail time to make a desired turn does not exceed 1 hour. Further away from the sun solar irradiance drops as $\frac{1}{r^2}$ which leads to decrease of radiation pressure force. However, even in this case performing a $180^o$ turn takes less than several days at a distance of 30 AU (Neptune's orbit) even for large area sails. Turn time strongly depends on the sail size, and grows as the sail size grows, as expected. This trend can be understood with the rapid increase of the mass moment of inertia (grows as $I_{zz} \propto L^4$), which is not compensated by a related increase in the torque (torque also grows with the sail size: $M_z \propto L$). Therefore, for large size sails a larger area of coverage by anomalously reflecting metasurfaces may be needed to achieve desired torques (in the example discussed total metasurface area is $4\ m^2$).

In conclusion, we have examined principles of metasurface design for efficient in-plane momentum transfer under ambient sunlight. Our design based on segmented tapered patches allows for an ultrawide band, $> 400nm$, polarization independent anomalous reflection (>60%) across a large fraction of the solar spectrum. Our simplistic principles of design yield in-plane momentum coupling coefficient of $C_{||} \simeq 0.65\ nN/W$ under normal incidence and $C_{||} \simeq 0.5\ nN/W$ at an oblique incidence. We anticipate that the performance may be further enhanced with employing sophisticated optimization techniques [45] and with the use of a broader set of materials. Analysis of an example sail roll control further illustrates practical utility of the designed structure. Beyond solar sailing such polarization independent ultrawide diffraction tailored to operate under ambient sunlight can find use in such applications as solar concentrators [46], solar thermal desalination [47], solar fuels [48], and radiative cooling [49].

**Acknowledgments**

Authors thank useful discussions with Haonan Ling, Pavel Shafirin, Ho-Ting Tung. A.R.D is thankful to discussions with Andrew Heaton, Daniel Tyler, James Liau and Grover Swartzlander. A.R.D. acknowledges the support from NASA ECF program (grant #80NSSC23K0076), and partial support from NASA grant #80NSSC21K0954 and the Aerospace Corporation University Partnerships Program (award #4400000443).

**Author contributions**

T.J.J. did all simulations and derivations. T.J.J and A.R.D wrote the manuscript and analyzed all data. A.R.D supervised the project.

**Competing interests**

The authors declare no competing interests.